\documentclass[pra,letterpaper,twocolumn,superscriptaddress,floatfix]{revtex4}
\usepackage{graphicx,psfrag,amsmath,amssymb,amsfonts,bbm,latexsym,color,dcolumn,bm}

\newcommand{\gsim}{\lower1.0ex\hbox{$\;\stackrel{\textstyle>}{\sim}\;$}}
\newcommand{\lsim}{\lower1.0ex\hbox{$\;\stackrel{\textstyle<}{\sim}\;$}}
\newcommand{\e}{\mathnormal{e}}

\definecolor{oucrimsonred}{rgb}{0.6, 0.0, 0.0}
\definecolor{persianblue}{rgb}{0.11, 0.22, 0.73}
\definecolor{forestgreen}{rgb}{0.13,0.35,0.13}
\newcommand{\BR}{\mbox{BR\,}}
\newcommand{\mmu}{m_{\mu}}

\newcommand{\ellp}{\ell^{\prime}}
\newcommand{\M}{\rm M}
\newcommand{\xM}{\mathnormal{x}_{\scriptscriptstyle{\rm M}}}
\newcommand{\xmu}{\mathnormal{x}_{\mu}}

\newcommand{\p}{\scriptscriptstyle +}
\newcommand{\m}{\scriptscriptstyle -}
\newcommand{\emmp}{e^{\m}\!\mu^{\p}}
\newcommand{\epem}{e^{\p}\!e^{\m}}
\newcommand{\be}{\begin{equation}}
\newcommand{\ee}{\end{equation}}
\newcommand{\bea}{\begin{eqnarray}}
\newcommand{\eea}{\end{eqnarray}}

\begin{document}
	
\title{Bounds on charged-lepton flavor violations via resonant scattering}

\author{Emidio Gabrielli}
\thanks{emidio.gabrielli@cern.ch}

\affiliation{\mbox{Dipartimento di Fisica, Universit\`a di Trieste, and INFN, Sezione di Trieste, Via Valerio 2, 34127 Trieste, Italy}}

\affiliation{\mbox{NICPB, R\"avala 10, Tallinn 10143, Estonia}}

\author{Barbara Mele}
\thanks{barbara.mele@roma1.infn.it}

\affiliation{\mbox{INFN, Sezione di Roma, Piazzale Aldo Moro 2, 00185 Roma, Italy}}

\author{Roberto Onofrio}
\thanks{onofrior@gmail.com}

\affiliation{\mbox{Dipartimento di Fisica e Astronomia ``Galileo Galilei'', Universit\`a  di Padova, 
		Via Marzolo 8, Padova 35131, Italy}}

\affiliation{\mbox{Department of Physics and Astronomy, Dartmouth College, 6127 Wilder Laboratory, 
		Hanover, NH 03755, USA}}

\begin{abstract}
We explore the possibility of probing flavor violations in the charged-lepton sector by means of high-luminosity lepton-photon 
and electron-muon collisions, by inverting initial and final states in a variety of decay channels presently used to bound such violations. 
In particular, we analyse the resonant lepton, $\gamma\, \ell \to \ellp$, and neutral-meson, $ \emmp \to \phi,\eta,\pi^0\!$, scattering channels, 
whose cross sections  are  critically dependent on the colliding-beams energy spread, being particularly demanding in the case of leptonic processes. 
For these processes, we compute  upper bounds to the cross-section corresponding to present limits on the {\it inverse} decay channel rates.
In order to circumvent the beam energy spread limitations we extend the analysis to processes in which a photon accompanies the resonance 
in the final state, compensating the off-shellness effects by radiative return. These processes might be studied at future facilities with moderate 
energies, in case lepton-photon and electron-muon collisions with sufficiently high luminosity  will be available.
\end{abstract}

\maketitle
\section{Introduction}

Processes with violations of the generational leptonic number play a crucial role to test the standard model and to collect hints towards 
its possible extensions. While neutrino oscillations have been evidenced, lepton flavor violations for charged particles (CLFV hereafter) 
have not yet been observed. Some contribution to CLFV is expected in the standard model incorporating massive and mixing neutrinos, 
but at a level beyond the detectability for any foreseeable future \cite{Marciano:1977wx,Petcov:1976ff,BWLee1,BWLee2,Langacker:1988up}. 
Extensions of the standard model such as supersymmetric models or grand unified theories instead provide ranges for 
CLFV rates  which can be of phenomenological interest ~\cite{IHLee1,IHLee2,Barbieri:1995tw,ArkaniHamed:1995fs,Kosmas:1993ch,Raidal:2008jk}. 
After earlier attempts to detect such effects in the muon decay channel $\mu \rightarrow e \gamma$~\cite{Brooks:1999pu}, the currently 
more stringent test is provided by the MEG experiment, which finds at 90\%C.L. a branching-ratio (BR) bound 
$ {\rm BR}(\mu \to e \gamma)< 4.2 \times 10^{-13}\,$~\cite{TheMEG:2016wtm,Adam:2013gfn}. 
A factor 10 sensitivity improvement is expected after the MEG II upgrade~\cite{Baldini:2018nnn}. 
Further improved constraints will be provided by muon electron conversion $\mu N \to e N$ experiments~\cite{Bernstein:2019fyh,Diociaiuti:2020yvo,Yucel:2021vir,Adamov:2018vin} and electron-ion colliders \cite{Cirigliano:2021img}.

In this letter we discuss a complementary method to constrain possible CLFV, 
obtained by time reversal from a typical two-body CLFV decay channel. For example,
in the $\mu \to e \gamma$ case, one can consider the inverse resonant production
of a muon by the fusion of an electron and a photon of appropriate energy,
in the  $ \gamma e \to \mu$ scattering.  This process takes  advantage from the possibility to control 
the beam intensities of electrons and photons. In principle, high intensity electron beams, as the one 
used in synchrotron radiation sources, and high intensity photon sources might provide 
a higher sensitivity, and allow, in the absence of detected events, for more stringent CLFV bounds. 
While we will see that, due to the extremely narrow linewidth of the process, the $\gamma e^{\m}  \to \mu^{\m} $ case 
is not viable, we will discuss in detail more promising processes with broader resonances as the ones involving the 
$\tau$ lepton and pseudoscalar/vector mesons.

As a possible way to cope with the critical limitations connected to the finite beam-energy spread,
 we  will then extend the above discussion by analyzing non-resonant channels derived by radiative return from the previous class of resonant processes. 
 
 The plan of the letter is as follows. In Section II, we discuss the production cross sections 
 for  resonant lepton ($\gamma\, \ell \to \ellp$) and neutral-meson ($ \emmp \to \phi,\eta,\pi^0\!$) scattering channels, including  
 beam energy-spread effects on which the results are critically dependent. Cross-section upper bounds are computed on the basis of present  
experimental limits on the corresponding {\it inverse} decay channels. In Section III, we go  beyond the leading-order resonant cross sections,
 and present analytical results for the corresponding radiative-return channels  assuming a lepton-flavor violating Lagrangian  in the effective 
 field theory (EFT) approach. Finally, in Section IV, we present our conclusions.

\section{CLFV processes as resonant phenomena}

The observation of CLFV events in resonant production may proceed under very strict kinematic conditions, as discussed 
 in the following by recalling quite general considerations. Let consider a generic resonant process induced by 
the scattering of the $a$ and $b$ states  $a+b \to R \to f$, where  $R$ is a resonance, and $f$ its decay final state. 
The $R$ production in $a\,b$ collisions, in the reference frame where  the $a$ and $b$  momenta are  parallel but opposite, 
    occurs   when the following 
condition is  fulfilled
\begin{equation}
E_a E_b \;(1+\beta_a\beta_b)= \frac{1}{2}(m_R^2-m_a^2-m_b^2)\, , 
\label{kinematic}
\end{equation}
with $\beta_{a,b}=p_{a,b}/E_{a,b}$, where $E_{a,b}$ and $p_{a,b}$ are the energies and corresponding (modulus of) momenta 
respectively, that for $E_{a,b} \gg m_{a,b}$ can be approximated as $E_a E_b \simeq (m_R^2-m_a^2-m_b^2)/4$.

In order for the $R$ resonance to go on-shell, 
one in general needs an excellent control on the energy spread of the colliding beams.
 The beam energy distribution can usually be described by  a Gaussian function  (see for instance \cite{Greco:2016izi})

\bea
G(E,\Delta E)&=&\frac{1}{\sqrt{2\pi} \Delta E} \exp\left[-\frac{(E-E_0)^2}{2 \Delta E^2}\right]\,,
\label{Gauss}
\eea
where $E_0$ is the energy for which the beam intensity is maximum and $\Delta E$ the beam energy spread. 
Then, the cross section which takes into account the effect of the  beam broadening is obtained as a convolution integral 
in the beam energy $E$ of the Breit-Wigner (BW) distribution of the resonance with the Gaussian distribution of the beams in 
Eq. (\ref{Gauss}). 

Before including possible beams energy spread, the resonant cross section, for center-of-mass energies $E$ 
close to the resonance rest energy $m_R$, and in the non-relativistic limit, is given by
\bea
\sigma(E)=\frac{(2J+1)\, {\BR}_{\! i} {\BR}_{\! f}}{(2S_a+1)(2S_b+1)}\frac{4\pi}{p^2}
\frac{\Gamma_R^2}{4(E-m_R)^2+\Gamma_R^2}\, ,
\label{sigmaBW}
\eea
where $J$ is the total angular momentum of the resonance, $S_{a(b)}$ the spin of the initial $a(b)$  state, ${\BR}_{\! i}$ 
and  ${\BR}_{\! f}$ are the  branching ratios of the resonance decays into the initial ($R \to a\,b$) and final ($R \to f$) state, 
 respectively, and $p$ is the $a,b$ momentum in the $a+b$ center-of-mass frame. 
 

\begin{figure}[t!]
\begin{center}
\includegraphics[width=0.95\columnwidth]{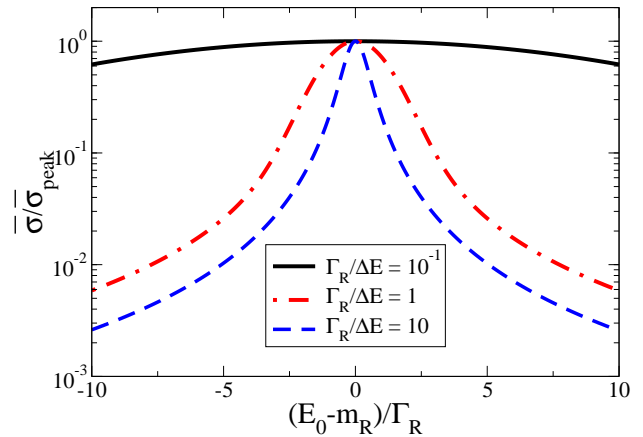}
\caption{\small Reduction of the integrated cross-section due to the mismatching between the peak of the beams energy distribution 
 and the resonance mass. The integrated cross-section normalized to the maximum of the cross-section $\bar{\sigma}_{\rm peak}$ is depicted versus the detuning factor 
 $(E_0-m_R)/\Gamma_R$, for three different values of the $\Gamma_R/\Delta E$ ratio, 
$10^{-1}$ (black, continuous), $1$(red, dot-dashed), and $10$ (blue, dotted).
\label{detuning} 
}
\label{Fig1}
\end{center}
\end{figure}


Then, the integrated or averaged cross-section over the beam energy spread at $E\simeq m_R$ is given by
\bea
\bar{\sigma}&\equiv &\int_{-\infty}^{\infty}
G(E,\Delta E) \sigma(E)dE\, .
\label{obscross0}
\eea
In the case of a narrow resonance $\Gamma_R\ll m_R$ the last term in 
Eq. (\ref{sigmaBW}) can be well approximated by a Dirac delta-function distribution, namely
\bea
\frac{\Gamma_R^2}{4(E-m_R)^2+\Gamma_R^2} \to \pi\frac{\Gamma_R}{2}\delta(E-m_R)\, .
\label{delta}
\eea
Then, in the narrow resonance approximation and for
$\Gamma_R/\Delta E \ll 1$, by means of Eq. (\ref{delta}), the cross section in Eq. (\ref{obscross0}) simplifies to
\bea
\bar{\sigma}
&\simeq& (2\pi)^{3/2} 
\frac{2J+1} {(2S_a+1)(2S_b+1)} 
\frac{{\BR}_{\! i} {\BR}_{\! f}}{2\,p_R^2}\, 
\frac{\Gamma_R}{\Delta E}\, ,
\label{obscross}
\eea
with, neglecting initial particle masses, $p_R\simeq m_R/2$. Likewise, if the resonance is broad, the energy distribution 
of the beams can be approximated with a Dirac delta-function distribution.

In a generic intermediate case, that is when $\Gamma_R$ and $\Delta E$ are of the same order, one needs to numerically integrate the 
observable resonant cross section. In this case, we evaluate the averaged cross section in Eq.(\ref{obscross0}) by retaining the whole energy 
dependence in Eq. (\ref{sigmaBW}), except for the momentum $p$, which has been replaced by its value $p\sim m_R/2$ at the resonant energy $E=m_R$, a condition valid in the $\Gamma_R/m_R\ll 1$ limit. In Fig.~1 we show the integrated resonant cross-section $\bar{\sigma}$ in this general case, 
normalized to its peak value -- that is relative to the case of an ideal tuning with a beam of negligible width -- as a function of the 
energy detuning relative to the process linewidth, for different values of the ratio $\Gamma_R/\Delta E$. Notice that, by construction, the 
ratio plotted in  Fig.~1 does not depend by the resonant mass $m_R$.  As expected, the suppression is negligible if the resonance is narrower 
that the spread of the energy beam, as evidenced in the $\Gamma_R/\Delta E =10^{-1}$ case. However, in the opposite case of 
$\Gamma_R/\Delta E=10$, the reduction is considerable, for instance about two orders of magnitude for a detuning $|E_0-m_R| = 5 \Gamma_R$.

\subsection{The $\gamma \ell \to \ellp$ processes}

We specialize here the considerations developed in the former section to the class of resonant FCNC processes in the charged lepton sector
\bea
\gamma \ell \to \ellp \, ,
\eea
for $\ell\neq \ellp$, with $\ell=e,\mu$ and $\ellp=\mu,\tau$.
By taking into account the beam-energy broadening effect in Eq.~(\ref{Gauss}), setting $E_0=m_{\ellp}$ and  $\Gamma=\Gamma_{\ellp}$, with 
$m_{\ellp}$ and $\Gamma_{\ellp}$ the $\ellp$ mass and width,  respectively,  and neglecting $m_{\ell}$, we obtain for the integrated resonant cross section
\bea
\bar{\sigma}(\gamma \ell \to \ellp)&\simeq &\frac{(2\pi)^{3/2}}{m_{\ellp}^2}\frac{\Gamma_{\ellp}}{\Delta E} \;{\BR}(\ellp \to \ell \gamma)\, .
\label{BWres}
\eea
where in Eq.~(\ref{obscross}) we have replaced ${\BR}_{\! i} \to {\BR}(\ellp \to \ell \gamma)$, and assumed 
${\BR}_{\! f}= 1$.

We analyze first the most promising $\ellp=\tau$ case, in particular the processes $\gamma e \to \tau$ and $\gamma\mu \to \tau$. 
By taking into account the $\tau$ total width  $\Gamma_{\tau}\simeq 2.27\times 10^{-3} \,{\rm eV}$ and
the present limits at 90\% C.L.  on the branching ratios for these processes, namely \cite{Aubert:2009ag}
\be{\rm BR}(\tau\to e \gamma)< 3.3\times 10^{-8} , \,
{\rm BR}(\tau\to \mu \gamma)< 4.4 \times 10^{-8},
\label{BRlimits-tau}
\ee
we get  the following upper bounds for the corresponding resonant cross sections
\bea
\bar{\sigma}(\gamma e \to \tau)&\lsim&1.4\left(\frac{{\rm 100\, keV}}{\Delta E}\right)~ {\rm ab}\, ,
    \\
\bar{\sigma}(\gamma \mu \to \tau)&\lsim&1.9 \left(\frac{{\rm 100\, keV}}{\Delta E}\right)~ {\rm ab}\, .
    \label{taugammabounds}
\eea
Hence, assuming a beam energy spread $\Delta E\sim{\rm 100\, keV}$ in  $\gamma e$ and $\gamma  \mu$ collisions, measurements of the 
$\gamma \ell \to \tau$ cross sections 
 of the order or smaller than $ 1 \,{\rm ab}$ are required to achieve sensitivities on the branching ratios for the CLFV $\tau\to e \gamma$ 
 and $\tau\to \mu \gamma$ decays  which are comparable or stronger than the corresponding upper limits in Eq.~(\ref{BRlimits-tau}).

It is also interesting to compare the latter cases with the more challenging resonant  $\gamma e \to \mu$ production.
With a muon lifetime of $\tau_{\mu} =2.2 \times 10^{-6}$ s (corresponding to a  total width  $\Gamma_{\mu}= 3.0 \times 10^{-10}$ eV),   
muon production would require an unrealistic beam energy spread. 
Moreover, the extremely constraining MEG bound on the $\mu \rightarrow e \gamma$ decay
($ {\rm BR}(\mu \to e \gamma)\lsim 4.2 \times 10^{-13}\,$ at 90\% C.L.~\cite{TheMEG:2016wtm})
implies an upper bound on the muon resonant cross section

\bea
\bar{\sigma}(\gamma e \to \mu)&\lsim&6.9\times 10^{-10}\left(\frac{{\rm 100\, keV}}{\Delta E}\right)~ {\rm ab}\, .
\label{mugammabound}
\eea
which is beyond any realistic experimental performance by many orders of magnitude. 

In preparation for the discussion presented in the next sections, it is useful to introduce an effective  Lagrangian responsible for 
the CLFV  $\ellp \to \ell \gamma$ decay. This effective Lagrangian will be written in terms of the lowest (gauge-invariant) 
dimensional local operator given by the magnetic-dipole like interactions 
\bea
    {\cal L}_{CLFV}&=&\sum_{\ell,\ellp} \frac{1}{2\Lambda_{\ellp 
    \!\ell}}
    \bar{\psi}_{\ell} \sigma^{\alpha\beta} \psi_{\ellp} ~ F_{\alpha\beta}+h.c.\, ,
    \label{Lmuegamma}
\eea
where $\Lambda_{\ellp\!\ell}$ is the associated mass effective scale, $\psi_{\ellp}$ and $\psi_{\ell}$ stand for the initial  and final heavier 
lepton fields respectively, $F_{\alpha\beta}=\partial_{\alpha} A_{\beta}-\partial_{\beta} A_{\alpha}$ is 
the usual electromagnetic tensor operator, with $A_{\alpha}$ the photon field, and $\sigma^{\alpha \beta}\equiv i/2[\gamma^{\alpha}, \gamma^{\beta}]$.
In a potential UV completion for the new physics (NP) scenario, the effective interaction in Eq.~(\ref{Lmuegamma}) could arise for instance at one loop. 
In this case, the effective $\Lambda_{\ellp \!\ell}$  scale is expected to be related to the mass and couplings of the NP scenario as
$1/\Lambda_{\ellp\!\ell}\sim g_{\ellp\!\ell}^2/(16 \pi^2 M_{\rm NP})$, where $M_{\rm NP}$ stand for the smallest mass relevant to new physics running 
in the loop, and $g_{\ellp\!\ell} \ll 1$ parametrizes  the corresponding dimensionless CLFV coupling.

The decay width for $\ellp\to \ell \gamma$  then becomes \cite{Gabrielli:2016cut}
\be
\Gamma(\ellp \to \ell \gamma)=\frac{m_{\ellp}^3}{8\pi \Lambda_{\ellp \!\ell}^2}
\left(1-r_{\ell}\right)^3\, ,
\label{Gammamuegamma}    
\ee
with $r_{\ell}=m_{\ell}^2/m_{\ellp}^2$.
For instance, in the case of the CLFV $\mu\to e \gamma$  decay, by using the upper bound  
${\BR}(\mu\to e \gamma)< 4.2 \times 10^{-13}$, from Eq.~(\ref{Gammamuegamma}) we obtain $\Lambda_{\mu e} \gsim 2 \times 10^{10} ~ {\rm TeV}$ 
that, for $M_{\rm NP}\!\sim 10$ TeV, would imply for the relevant CLFV coupling  $g_{\mu e}\!\lsim 3 \times 10^{-4}$.

\subsection{The $ \emmp  \to \phi,\eta,\pi^0$ processes}

The severe bounds on the cross section for  inverse CLFV processes like $e (\mu) \gamma \to \tau$ or $e \gamma \to \mu$ are
due to both the extremely narrow width of the final state, and the strong experimental bound on the related branching ratios, especially 
in the $\mu\to e \gamma$ case. These two conditions are not satisfied for processes 
involving broader resonant states and less stringent experimental constraints on CLFV branching ratios (see \cite{Nussinov:2000nm} for 
model-independent bounds based on unitarity).

In order to consider less constrained setups, we then extend our discussion
to CLFV inverse processes involving neutral mesons, hence replacing 
$\gamma \ell$ initial states by opposite-charge leptons of  different flavor.
 In particular, we will analyze  the resonant CLFV production of a neutral meson $\M$, via the channel $e^\pm \mu^\mp \to \M\,$.
Among many possibilities, we restrict the choice to the pseudoscalar mesons such as the neutral pion $\pi^0$ and the $\eta$, as well as the vector meson $\phi$. 
The considered mesons have widths much smaller than their mass, even in the broader $\phi$ case  ($\Gamma_\phi/m_\phi \simeq 0.5 \%$), and 
 Eq.~(\ref{obscross}) still gives a proper description for the effective $e \mu \to \M$ cross sections.
 
The resonant $\M=\pi^0\!,\eta,\phi$ cross section averaged over the beam energy spread is then, analogously to  Eq.~(\ref{BWres}) but retaining the exact muon mass ($\mmu$) dependence, 
\bea
\bar{\sigma}({\emmp \to \M})\simeq \frac{C_{\M}(2\pi)^{3/2}}{m_{\M}^2(1-r_{\mu})^2}  \;\frac{\Gamma_{\M}}{\Delta E}  \,\BR(\M \to \e^- \mu^+) ,\;\;\;\;\;
\label{BWresmeson}
\eea
with $m_{\M}$ and $\Gamma_{\M}$ the $\M$ meson  mass and width,  
respectively, while the coefficient $C_{\M}$ ($C_{\pi^0\!,\eta}=1/2$, and  $C_{\phi}=3/2$) accounts for the $\M$ spin degeneracy, and 
$r_{\mu}=\mmu^2/m^2_{\M}$.

By using the corresponding  meson widths, $\Gamma_{\phi}\simeq 4.25~ {\rm MeV}$,
$\Gamma_{\eta}\simeq1.31~ {\rm keV}$,  and 
$\Gamma_{\pi^0}\simeq7.72~ {\rm eV}$, and the present experimental upper limits on the CLFV branching ratios at 90\% C.L.,
\bea
{\rm BR}(\phi \to e \,\mu) &\lsim& 2 \times 10^{-6},~~~~~~
    \text{\cite{Achasov:2009en}}\nonumber \\
{\rm BR}(\eta \to e \,\mu) &\lsim& 6\times 10^{-6},~~~~~~
  \text{\cite{White:1995jc}}\nonumber \\
{\rm BR}(\pi^0 \!\to e \,\mu) &\lsim& 3.6\times 10^{-10},~~~
\text{\cite{Abouzaid:2007aa}}
\label{expboundsmeson}
\eea
we obtain from Eq. (\ref{BWresmeson}) the following upper bounds on the cross sections 
\bea
\bar{\sigma}(\emmp \to \phi)&\lsim& 3.8\times 10^2 \left(\frac{{\rm 100\, keV}}{\Delta E}\right) {\rm nb}\, , \\
\bar{\sigma}(\emmp \to \eta)&\lsim& 4.3\times 10^2 \left(\frac{{\rm 100\,keV}}{\Delta E}\right) {\rm pb}\, ,  \\ 
\bar{\sigma}(\emmp \to \pi^0)&\lsim& 1.5 \times 10^{-2}\!\left(\frac{{\rm 100 \,keV}}{\Delta E}\right) {\rm pb}\, .
\label{mesonbounds}
\eea
In the latter equations, which include only one charge combination for the colliding particles, we use half of  the BR's experimental upper limits in Eqs.~(\ref{expboundsmeson}).

Electron-muon collisions in the  energy range needed for the latter processes are indeed presently under consideration. 
The fixed target experiment MUonE plans to explore electron-muon scattering at $\sqrt{s}\simeq $ 400 MeV~\cite{Banerjee:2020tdt}. 
By running a similar experiment at lower $\sqrt{s}$, one in principle could cover the CLFV channel $\emmp \rightarrow \pi^0$, which 
has been investigated in the decay mode \cite{Krolak:1994aj,Appel:2000tc}, but  has an expected $ \pi^0 \!\to e \,\mu$  partial width  
which is less than about $ 3 \times 10^{-9}$ eV. With a modest boost in the energy with respect to MUonE, one could also reach the 
significantly broader $\eta$ resonance.

Notice that muon beams are expected to have a relative energy spread much smaller than their electron counterpart, due to the 
low impact of bremsstrahlung and synchrotron radiation, which is estimated to be at the Higgs-boson peak of the order 
of $\Delta E/E \simeq 10^{-5}$ \cite{Ankenbrandt:1999cta,Raja:1998ip}.

A different possibility might be provided by non fixed-target electron-muon setups.  For instance, in order to go on the $\eta$
resonance, assuming an electron beam with $E_e$=50 MeV, one would need a muon beam with $E_{\mu}=$1.4~GeV. This combination could 
match the use of a high intensity electron beam from a van der Graaf accelerator (with order of $\mu$A currents) and a relatively high-energy muon 
beam  with beam energy spread $\Delta E \simeq $ 15~keV, with a comparatively negligible electron beam energy spread. 

In analogy to the purely leptonic channels, it  is convenient to parametrize the effective CLFV coupling between the generic meson $\M$ (with $\M=\pi^0,\eta$ for the case of scalar mesons, and $\M=\phi$ for the vector meson $\phi$) and the leptons.
 We assume the effective Lagrangian to be dominated by lowest dimensional operators of dimension 4. 
 In particular, for the neutral scalar mesons $\M=\eta,\pi^0$, we parametrize it with operators of scalar Yukawa-type interaction
\be
{\cal L}^{S}_{\rm CLFV}=\left(Y^L_{\mu e}\bar{\psi}_{\mu}P_L \psi_e+
Y^R_{\mu e} \bar{\psi}_{\mu}P_R \psi_e\right) \varphi_M + h.c.\, ,~
\label{effectiveLmeson}    
\ee
where we assume the most general parity-violating couplings $Y^{L,R}_{\mu e}$ to the scalar meson $\M$ induced by some new physics, with the chiral 
projectors defined as $P_{L/R}=(1 \mp \gamma_5)/2$, and $\varphi$ stands generically for the scalar field associated to the scalar meson $\M$. On the other hand, for the vector meson $\phi$, we can parametrize the corresponding interaction as
\be
{\cal L}^{V}_{\rm CLFV}=\left(Y^L_{\mu e}\bar{\psi}_{\mu}\gamma^{\alpha} P_L \psi_e+
Y^R_{\mu e} \bar{\psi}_{\mu}\gamma^{\alpha} P_R \psi_e\right) V_{\alpha} + h.c.\, ,~
\label{effectiveLPhi}    
\ee
where $V_{\alpha}$ stands for the massive spin-1 field associated to the $\phi$ meson.
To simplify the notations the same symbols for the couplings as in Eq.(20) have been adopted.

The effective couplings appearing in Eqs. (\ref{effectiveLmeson},\ref{effectiveLPhi}) can originate at the fundamental level of quark and lepton interactions, via CLFV dimension 6 operators induced for instance by the exchange of some heavy new physics particle like
  \bea
      {\cal L}_{\rm eff}\sim \frac{1}{\Lambda^2} [\bar q \Gamma_i q][\bar \mu \Gamma_i e]+h.c. \, ,
      \label{dim6}
  \eea
  where $q=u,d,s$ are the light quark fields,
  and $\Gamma_i$ generically indicate matrices of the Clifford basis (contraction over Lorentz indices is understood). For instance, in the case of a vectorial exchange, $\Gamma_i=\gamma_{\mu}$ or $\Gamma_i=\gamma_{\mu}\gamma_5$, one can easily relate the scale $\Lambda$ to the effective Yukawa coupling $Y^{L,R}_{\mu e}$ by making use of Lorentz invariance to compute the hadronic matrix element of
the  $\bar{q} \gamma_{\mu}\gamma_5 q$ operator between the vacuum and the meson states. In particular, for the $\pi^0$ we have
\bea
Y^{L,R}_{\mu e}\sim \frac{f_{\pi}m_{\mu} }{\Lambda^2}\, ,
\eea
with $f_{\pi}$ the pion decay constant, that for a new physics scale of the order of $\Lambda\simeq 1~  {\rm TeV}$ yields $Y^{L,R}_{\mu e}\sim 10^{-8}$.
Analogous results can be obtained for the $\eta$ and $\phi$ mesons, by applying the same considerations. Then, due to the short-distance nature of the quark-lepton interaction in Eq. (\ref{dim6}), the validity of the effective meson interactions in Eqs. (\ref{effectiveLmeson},\ref{effectiveLPhi}), is up to energies of the order of the scale $\Lambda$.
  
By using the effective Lagrangians in Eqs.~(\ref{effectiveLmeson},\ref{effectiveLPhi}), and by neglecting contributions proportional to $m_e$, 
we obtain, for the total width for the CLFV meson decay $\M \to \e^-\!\mu^+$

\bea
\Gamma(\M \to \e^-\!\mu^+)&=&\frac{S_{\M}|Y_{\mu e}|^2 \;m_{\M}}{8\pi}  \left(1-r_{\mu}\right)^2\, , 
\label{meson-width}
\eea
where $|Y_{\mu e}|^2\equiv \left(|Y^L_{\mu e}|^2+|Y^R_{\mu e}|^2\right)/2$, with spin factors 
 $S_{\eta,\pi^0}=1$, $S_{\phi}=\left(2+r_{\mu}\right)/3$.

By using the bounds in Eq. (\ref{expboundsmeson}) we can derive the upper limits at 90\% C.L.  on the corresponding CLFV couplings $Y_{\mu e}$
associated to the meson $\M$. In particular, we have
\bea
|Y_{\mu e}|&<&4.0\times 10^{-4}\, , ~~~ \M=\phi
\nonumber \\
|Y_{\mu e}|&<&1.4\times 10^{-5}\, , ~~~ \M=\eta
\nonumber \\
|Y_{\mu e}|&<&4.2\times 10^{-8}\, , ~~~ \M=\pi^0\, .
\eea


 \begin{figure}[t]
\begin{center}
\includegraphics[width=0.90\columnwidth]{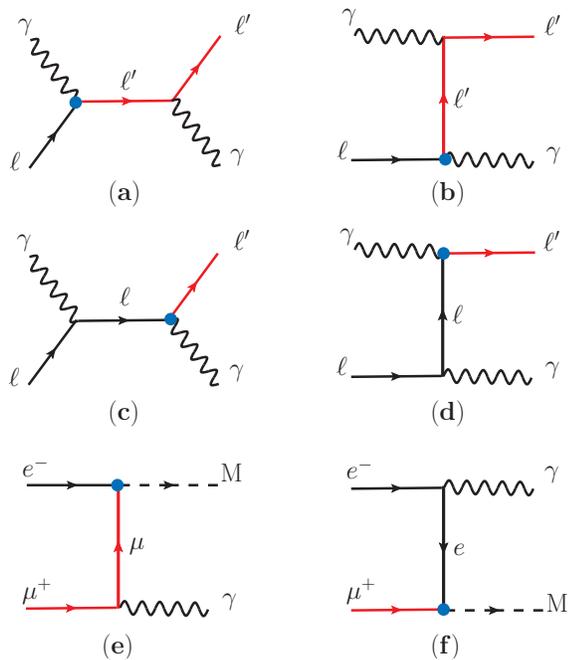}
\caption{\small Tree-level Feynman diagrams for the radiative return processes  $\gamma \ell \to \gamma \ellp$ with 
$\ell \neq \ellp$ (a-d),  and $\emmp \to \M \gamma\,$ (e,f) with  $\M=\phi,\eta,\pi^0$.
The bubble vertex represents the proper effective CLFV interaction.}
\label{Fig2}
\end{center}
\end{figure}


\section{Radiative return effects}

It is clear from Eq.~(\ref{BWres}) that the $\Gamma_{\mu}/\Delta E$ ratio is a crucial factor affecting  the resonant  cross section, which  
 can be particularly demanding in the case of a very narrow resonance, most notably in the  leptonic cases considered above. 
This limitation can be circumvented by considering the radiative photon emission (giving rise, for instance, to the $\gamma e \rightarrow \gamma \mu$ 
process in  the $\gamma e \rightarrow \mu$ case),  an effect known as  {\em radiative return}~\cite{Yennie:1961ad,Barger:1996jm}. 
This  has been mainly studied for the production of neutral resonant states (like $J/\Psi$, $Z^0$,  Higgs boson) in $\epem$ annihilation 
out of the resonant region, where it played a major role in the discovery of the $J/\Psi$.
In these cases, its contribution can be reabsorbed into the  initial-state-radiation (ISR) effects, which also include higher-order QED corrections and their exponentiation \cite{Yennie:1961ad,Kuraev:1985hb,Nicrosini:1986sm,Jadach:2000ir}.

In the next two subsections, we discuss  possible advantages  of the radiative-return channels for CLFV processes, distinguishing between the 
production of leptonic and mesonic final states.

\subsection{The $\gamma \ell \to \gamma \ellp$ processes}

An extra photon emission in the $\gamma \ell \to \ellp$ channel, where the resonance is a charged state, will involve both the initial and final states.
Since the operator inducing the CLFV vertex for the $ \ell \gamma \to \ellp$ transition in Eq.~(\ref{Lmuegamma}) is a (dimension-5) magnetic-dipole 
operator, the $\gamma \ell \to \gamma \ellp$ cross section will behave in the high energy limit in a dramatically different way with respect to the usual renormalizable interactions induced by dimension-4 operators. Indeed, we will see that it will tend to a constant in the asymptotic energy limit, against the usual $1/s$ cross-section behaviour of renormalizable interactions. 

Here we provide an estimation of the upper bound of the CLFV 
$\gamma \ell \rightarrow \gamma \ellp$ cross sections arising  from radiative return effects,  and compare them with the corresponding resonant processes 
cross sections discussed in section II.

Let us now consider the CLFV scattering process,
induced by the Lagrangian in Eq.~(\ref{Lmuegamma}),
\bea
\gamma(q_1)~e(p_1) \to~ \gamma (q_2)~ \mu (p_2)\,,
\label{gegmu}
\eea
 where $p_{1,2}$ and $q_{1,2}$ are the relevant 4-momenta. The corresponding Feynman diagrams are shown in Fig.~2, panels (a-d). 
Generalizations to CLFV radiative transitions $\ell \to \ellp$ involving other lepton flavors are straightforward.
We obtain for the differential cross section for the process in Eq.~(\ref{gegmu}) 
\bea
\frac{d \sigma}{d t}(\gamma e \to \gamma \mu)&\simeq&
\frac{\alpha}{2\Lambda_{\mu e}^2}\frac{F(s,t)}{
  (s-\mmu^2)^2+\Gamma_{\mu}^2\mmu^2}\, ,
\label{dsigmarad}
\eea
where $\alpha$ is the fine structure constant evaluated at the $\mmu$  scale, and $s=(p_1+q_1)^2$, $t=(q_1-p_2)^2$ 
are the Mandelstam variables. The function $F(s,t)$ is given by
\bea
& & F(s,t)=\frac{s+t-\mmu^2}{ s^3(t-\mmu^2)^2 (t-m_e^2)} \times \nonumber \\
& & \Big[6s^3t^3-12\mmu^2s^2t^2(s+t)+\mmu^4st(7s^2+24st+7t^2)-\nonumber\\
& &\mmu^6(s+t)(s^2+13st+t^2)+12 \mmu^8st-\mmu^{10}(s + t)\Big)\Big]\, \nonumber,
\label{F}
\eea
where the electron mass $m_e$ is retained only in the denominator of the $t$-channel propagator, as needed to regularize the collinear divergencies. 
The exact expression for the function $F(s,t)$ with the full mass dependence is reported in Appendix A.
For a soft final photon, in proximity of the kinematic threshold given by   $s\sim \mmu^2$, the resonant term in Eq.~(\ref{dsigmarad}) is 
regularized by the muon width $\Gamma_{\mu}$.


\begin{figure*}[t!]
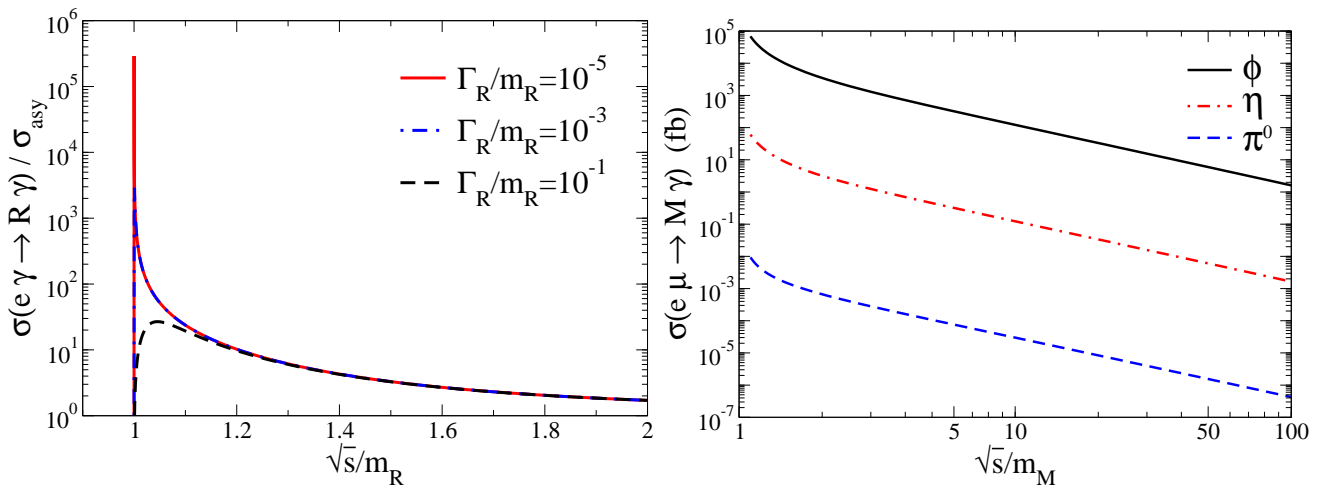

\begin{center}
  \includegraphics[width=0.48\textwidth,clip=true]{CLFVFig3a.eps}
  \includegraphics[width=0.48\textwidth,clip=true]{CLFVFig3b.eps}
  \caption{\small 
(Left) Cross section $\sigma(e\gamma \to R \gamma)$ versus $\sqrt{s}/m_R$, for the 
 CLFV production of a fermion resonance $R$,  normalized to the corresponding 
 asymptotic cross section $\sigma_{\rm asy}(e\gamma \to R \gamma)$ at large $\sqrt{s}\gg m_{\rm R}$, for $m_R=\mmu$,  and 
 $\Gamma_{R}/m_R=10^{-5}$ (red, continuous), $10^{-3}$ (blue, dot-dashed),  $10^{-1}$ (black, dotted).
(Right) Upper bounds on total cross sections   for the $\M$ meson  production $\emmp\to \M \gamma$, versus  $\sqrt{s}/m_{\M}$, for  
$\M=\phi$ (black, continuous), $\M=\eta$ (red, dot-dashed), and  $\M=\pi^0$ (blue, dotted). A cut $E_\gamma \gsim m_{\M}/10$ is applied.
  }
\label{Fig3} 
\end{center}
\end{figure*}


We first analyze this cross section in the asymptotic regime of center-of-mass energies  $\sqrt{s}\gg m_{\mu}$, 
where  $\mmu$ (and hence $m_e$) can be neglected. By taking the massless limit in Eq.~(\ref{dsigmarad}), we obtain 
\be
\frac{d \sigma_{\rm asy}}{d t}(\gamma e \to \gamma \mu)\simeq\frac{3\alpha}{\Lambda^2_{\mu e}}~\frac{s+t}{s^2}\, , 
\label{dsigmamue}
\ee
and by integrating Eq.~(\ref{dsigmamue}) over the entire phase space  ($-s \le t \le 0$), we get for the {\it asymptotic} total cross section
\be
\sigma_{\rm asy}(\gamma e \to \gamma \mu)\simeq \frac{3\alpha}{2\Lambda_{\mu e}^2}\left(1+{\cal O}(m_{\mu}^2/s)\right).
\label{crossrad0}
\ee

The exact result, reported in Appendix A, allows to check that no powerlike  singularities as from terms $\sim~\!\!\!1/t^2$, potentially 
modifying the expectations in Eq.~(\ref{crossrad0}) by contributions of the same order,  arise in the muon and electron massless limit. 
The usual enhancement terms $\sim~\!\!\! \log{(s/m^2_\ell)}$, arising from almost collinear, chirally suppressed, kinematic configurations, are 
all included in the last term of Eq.~(\ref{crossrad0}). Therefore, the leading contribution to the total cross section tends to a constant at 
high energies $\sqrt{s}\gg \mmu$. This is simply due to dimensional reasons, being the effective operator in Eq.~(\ref{Lmuegamma}) of dimension 5. 
However, it should be kept in mind that the interaction in Eq.~(\ref{Lmuegamma}) is an effective low energy coupling, 
which is valid up to characteristic scattering energies  of the order $\sqrt{s} \lsim {\cal O}(\Lambda_{\mu e})$, 
above which a UV completion of the theory should be taken into account.

By using the $\mu\to e \gamma$ decay width  in Eq.~(\ref{Gammamuegamma}), we can rewrite the total cross 
section in  Eq.~(\ref{crossrad0}) in terms of the branching ratio ${\rm BR}(\mu\to e \gamma)$ as
\be
\sigma_{\rm asy}(\gamma e \to \gamma \mu)\simeq 12 \pi \alpha \;\frac{\Gamma_{\mu}}{m^3_{\mu}}\; {\rm BR}(\mu\to e \gamma)\, ,
\label{cross-as}
\ee
which shows the insensitivity of the radiative return rate in the asymptotic regime to the beam energy spread. 
By using the existing experimental bound on   $\BR(\mu\to e \gamma)$,  we get an upper limit on the asymptotic radiative cross section 
\be
\sigma_{\rm asy}(\gamma e \to \gamma \mu)\lsim 1.15 \times 10^{-14}~ {\rm ab} \, ,
\label{bound-cross-as}
\ee
that is anyhow  too tiny to give measurable effects.

Independently from the ${\BR}(\mu \to e \gamma)$ bound, the CLFV resonant $e \gamma \to \mu$ cross section, as obtained from 
Eq.~(\ref{BWres}), is in general dominant with respect to the corresponding non-resonant $\gamma e \to \gamma \mu$ one.
In the asymptotic  regime of Eq.~(\ref{crossrad0}),  their ratio is given by
\be
\frac{\sigma_{\rm asy}(\gamma e \to \gamma \mu)}{\bar{\sigma}(e \gamma \to \mu)} \simeq \frac{6\alpha}{\sqrt{2\pi}}
\frac{\Delta E}{m_{\mu}}\, ,~
\label{crossratio}
\ee
with the two cross sections becoming comparable for a beam energy spread $\Delta E \sim 57\, m_{\mu}$.
Although neither  cross sections are  realistically measurable, in principle, for a typical  beam energy spread of 
$\Delta E \sim 100 \,{\rm keV}$, running on the $\mmu$ pole would be more advantageous than looking
 at the radiative process with non-tuned collider facilities outside the resonant region. 

Analogous results can be obtained for the more promising $\gamma e \to \gamma \tau$ and $\gamma \mu \to \gamma \tau$ 
processes, by properly replacing in Eq. (\ref{cross-as}) the muon width and mass, and the ${\rm BR}(\mu\to e \gamma)$ bound with the $\tau$ corresponding quantities. In particular, by using the upper limits in Eq. (\ref{BRlimits-tau}), we obtain
\bea
\sigma_{\rm asy}(\gamma e \to  \gamma \tau) &\lsim& 1.4\times 10^{-6} ~{\rm ab}\, ,
\nonumber \\
\sigma_{\rm asy}(\gamma \mu \to \gamma \tau) &\lsim& 1.9\times 10^{-6}~ {\rm ab} \, .
\label{bound-cross-as-tau}
\eea
By extending   Eq. (\ref{crossratio})  to the $\tau$ production, the radiative mode is confirmed to be sub-leading with respect to 
the resonant one, unless the energy spread becomes unrealistically large and of the order of at least $\Delta E \sim 100 \,{\rm GeV}$.

On the other hand, in energy regions close to the resonance threshold, implying emissions of soft IR photons, the radiative $\gamma \ell \to \gamma \ellp$ 
process can  present a large enhancement inversely proportional to the resonance width. 
The exact total cross section, normalized to the asymptotic cross section, versus the center-of-mass energy $\sqrt{s}$  is shown in  Fig. \ref{Fig3} (left plot) 
for a  charged-lepton resonance $R$.  For illustrative purposes we show three representative cases for the relative resonance width,  $\Gamma_{R}/m_R=10^{-5},10^{-3},10^{-1}$, valid for $m_R=\mmu$. 
The cross section peaks at more than $10^5$  its asymptotic value for the case of $\Gamma_{R}/m_R\simeq10^{-5}$.
Nevertheless, the potential gain in the radiative process is expected to be smeared out when convoluted with  realistic beam energy spreads. 
The shapes of the curves in the left plot of Fig. \ref{Fig3} are almost independent on the mass $m_R$ in the resonant region, since this explicit 
dependence slightly affects the overall normalization alone.

For completeness, we provide  an analytical approximated expression of the 
$\gamma e \to \gamma \mu$ total cross section valid
for energies close to the threshold region $\sqrt s\simeq m_{\mu}$ 
(results can be  generalized to different flavor radiative processes in a straightforward way).
In particular, following the results in Appendix A, the total cross section near the resonant region can be written as
\bea
\sigma_{\rm thr}(\gamma e \to \gamma \mu) &\simeq& 8\pi \alpha \;C^{e\mu}_F \frac{\Gamma_{\mu}}{\mmu} \;{\rm BR}(\mu\to e \gamma)\times \nonumber \\ 
&&[\;\frac{s-\mmu^2}{(s-m_{\mu}^2)^2+\Gamma_{\mu}^2\mmu^2} \;],~~
\label{rad-distr-peak}
\eea
where $C^{e\mu}_F\simeq 8.7$ (see Appendix A) accounts for an average on the integrated angular distribution near the $\sqrt{s}$ peak value. 
The $s$-distribution in square brackets  has a maximum $\sigma_{\rm peak}(\gamma e \to \gamma \mu)$ at $s=\bar{s}$ with $\bar{s}=\mmu^2(1+\Gamma_{\mu}/\mmu)$, and vanishes at the 
threshold $s\to \mmu^2$. The peak value for the cross section is
\bea
\sigma_{\rm peak}(\gamma e \to \gamma \mu)\simeq\frac{4 \pi \alpha\, C^{e \mu}_F}{\mmu^2}\;
\BR(\mu\to e \gamma)\,  \lsim \,  12~ {\rm fb}\, .
\nonumber
\label{sigmaradmax}
\eea

 The corresponding peak values for $\tau$ production cross sections are 
\bea
&&\sigma_{\rm peak}(\gamma e \to  \gamma \tau)\lsim \,  5.3 ~{\rm pb}\, ,
\nonumber \\
&&\sigma_{\rm peak}(\gamma \mu \to \gamma \tau)\lsim \, 1.8 ~ {\rm pb}\, .
\label{sigmaradmax-tau}
\eea
If we compare the above result with the  BW cross section in Eq.~(\ref{sigmaBW}), we find for the ratio of the radiative versus the non-radiative peak values 
\be
\frac{\sigma_{\rm peak}(\gamma e \to \gamma \mu)}{\sigma_{\rm peak}(\gamma e \to \mu)}\simeq \frac{\alpha}{2} C^{e\mu}_F \simeq 0.032\, , 
\label{ratiopeak}
\ee
where $\sigma_{\rm peak}$ is the corresponding BW distribution in Eq. (\ref{sigmaBW}) evaluated at the resonant energy
$E=m_R$.

Then, for  the {\it averaged}  (according to the integral in Eq.~(\ref{obscross0}) with $E_0\sim \mmu$) radiative cross sections near the 
peak region, we obtain
\be
\bar{\sigma}_{\rm thr}(\gamma e \to \gamma \mu)\simeq \frac{\alpha \,C^{e\mu}_F (2\pi)^{3/2}}{\mmu^2} 
\frac{\Gamma_{\mu}}{\Delta E}\;\BR(\mu \to e\gamma),
\label{sigmathrave}
\ee
which is similar, apart from a numerical factor  ($\sim\alpha$), to the result for the resonant BW cross section in Eq. (\ref{BWres}). 
In deriving Eq. (\ref{sigmathrave}) the approximation of the relativistic version of the distribution in Eq. (\ref{delta}) for a narrow width has 
been used against the energy $E$ convolution integral in Eq. (\ref{obscross0}), namely $1/((s-m_R^2)^2+\Gamma^2 m_R^2)\to \pi/(2 \Gamma m_R^2) \delta(E-m_R)$, while the rest of the $s$ function appearing in Eq.(\ref{rad-distr-peak}) has been evaluated at its maximum value, that is for $s=\bar{s}$. Then the ratio of the convoluted (by beam energy spread effects) radiative cross section close to the threshold and the corresponding one for the BW production in Eq.~(\ref{BWres}) yields
\be
\frac{\bar{\sigma}_{\rm thr}(\gamma e \to \gamma \mu)}{\bar{\sigma}(\gamma e \to \mu)}\simeq \alpha C^{e\mu}_F,
\ee
which is approximatively twice the ratio of the unconvoluted peak cross sections in Eq.~(\ref{ratiopeak}).

\subsection{The $ e^- \mu^+  \to (\phi,\eta,\pi^0)\,\gamma$ processes}

We consider now the radiative return process applied to the resonant meson production $\emmp \to \M$ \bea
e^-(p_1) \mu^+(p_2) \to \M(q_1) \gamma(q_2)\, ,
\label{mesonrad}
\eea
where $\M$ indicates a neutral meson, for the particular cases $\M=\phi,\eta,\pi^0$. 
As shown in the corresponding Feynman diagrams in Fig.~\ref{Fig2} (e,f), the photon emission is uniquely due to  initial state radiation.


\begin{table*}[t]
\begin{center}
\begin{tabular}{|l|c|c|c|c|}
\hline
CLFV Process                & $E_a \times E_b ({\mathrm{GeV^2}})$      & $\Gamma_{\mathrm{tot}} ({\mathrm{MeV}})$ & BR$_{\rm exp}^{\rm max}$           &  $\bar{\sigma}^{\rm max}_{(\Delta E=100~{\mathrm{keV}})}  ({\mathrm{fb}})$  \\
\hline 
\hline
$\gamma \mu  \to \tau$   & $7.9 \times 10^{-1}$                                                      &  $2.27 \times 10^{-9}$                                    &  $4.4 \times 10^{-8}$   \cite{Aubert:2009ag}   &  $1.9 \times 10^{-3}$                 \\
$\gamma  e \to \tau$       & $7.9 \times 10^{-1}$                                                        &  $2.27 \times 10^{-9}$                                    &  $3.3 \times 10^{-8}$  \cite{Aubert:2009ag}   &  $1.4 \times 10^{-3}$                 \\
$\gamma e \to \mu$       & $2.8 \times 10^{-3}$                                 &  $~3.00 \times 10^{-16}$                                  &  $4.2 \times 10^{-13}$ \cite{TheMEG:2016wtm}    &  $~6.9 \times 10^{-13}$               \\
\hline
 $\emmp \to \phi$       &  $2.6 \times 10^{-1}$                                                      &  $4.25$                                                       &  $2.0 \times 10^{-6}$   \cite{Achasov:2009en}    &   $3.8 \times 10^{8}$                  \\
$\emmp \to \eta$        &  $7.2 \times 10^{-2}$                               &  $1.31 \times 10^{-3}$                                  &  $6.0 \times 10^{-6}$   \cite{White:1995jc}    &   $4.3 \times 10^{5}$                      \\
$\emmp \to \pi^0$      &  $3.5 \times 10^{-3} $                                &  $7.72 \times 10^{-6}$                                    &  $~3.6 \times 10^{-10}$ \!\cite{Abouzaid:2007aa}   &  $1.5\times 10^1$                   \\
\hline
\end{tabular}
\caption{Summary of the inverse CFLV processes discussed in the text, with leptons and mesons in the final state. 
The columns report, for each process, the energy product from Eq.~(\ref{kinematic}), the total resonance line-width, the current bound on the BR of 
the CLFV direct decay mode, and the corresponding bound on the integrated cross section for the inverse-decay scattering process, for  $\sqrt{s}$ tuned at the resonance with a beam energy spread $\Delta E$ of $100~ {\rm keV}$.
All bounds on the CLFV branching ratios are at the $90 \%$ confidence level. In the last row the product $E_a \times E_b$ corresponds to the frame in which one of the initial particles is at rest. } 
\end{center}
\label{table1}
\end{table*}


We keep the full $\mmu$ dependence, and the $m_e$ dependence only in the denominator of the electron 
propagator in order to regularize the collinear divergencies. By using the effective Lagrangians in 
Eqs. (\ref{effectiveLmeson},\ref{effectiveLPhi}), the differential cross section for the radiative return process in Eq.~(\ref{mesonrad}) becomes\\

\bea
\frac{d \sigma}{dt}(\emmp \to \M\gamma)\,=\,\frac{2\pi \alpha \,
  {\rm BR}(\M \to \e^- \mu^+)}{(1-\mathnormal{x}_{\mu})^2(1-\mathnormal{r}_{\mu})^2\mathnormal{S}_M}\;\frac{\Gamma_{\M}}{\it m_{\M}} {\mathnormal F}_{\M}(t)~~
\label{dsigma-mesonrad}
\eea
where
\bea
F_{\eta,\pi^0}(t)&=&\frac{\xM^2+1-2\xmu}{(t-\mmu^2)(u-m_e^2)}+\frac{2\mmu^2(1-\xM)
  (\xM-\xmu)}{(t-\mmu^2)^2(u-m_e^2)}\, ,
\nonumber\\
F_{\phi}(t)&=&\frac{2-5\xmu+2\xM(1+\xmu)+r_{\mu}(1-2\xmu)}{(t-m^2_{\mu})(u-m^2_e)}
\nonumber\\
&+&2\mmu^2\frac{\left(2\xM-\xmu(1+r_{\mu})\right)(1-\xM)}{(t-\mmu^2)^2(u-m_e^2)}
\nonumber\\
&+&\frac{3(1-\xM)-\xmu}{s(u-m_e^2)}+\frac{(t-\mmu^2)(3-5\xM)}{s^2(u-m_e^2)(1-\xM)}\, ,
\label{Fscalar}
\eea
with $x_a\equiv m^2_a/s$, $s=(p_1+p_2)^2$, $t=(p_1-q_1)^2$ and $u=(p_1-q_2)^2$.
By integrating over $t$, we get for the total cross section
\bea
\sigma(\emmp \to \M \gamma)\,=\,\frac{2\pi \alpha\,
  {\rm BR}(\M \to \e^- \mu^+)}{(1-\xmu)^2 \mathnormal{s}(1-\mathnormal{r}_{\mu})^2  \mathnormal{S}_M}\frac{\Gamma_{\M}}{\it m_{\M}} \mathnormal{\rho}_M(\xM,\xmu)\, , ~~
\eea
where the ISR effect can be factorized by the function $\rho_{\M}(\xM,\xmu)$ given by
\bea
&&\!\!\!\!\rho_{\eta,\pi^0}(\xM,\xmu)=\frac{2}{(1-\xM)}\Big[(\xM-\xmu)(\xmu-1)+
\nonumber\\
&&(1+\xM^2-2\xmu-2\xM\xmu+2\xmu^2)L_1
\Big]\, ,
\nonumber\\
\nonumber\\
&&\!\!\!\!\rho_{\phi}(\xM,\xmu)=
\frac{1}{(1-\xM)}\Big[2L_1\big(2-5\xmu+2\xM(1+\xmu)+
\nonumber\\
&& r_{\mu}(1-2\xmu)\big)-2(1-\xmu(1-2L_1)\big)(2\xM-\xmu(1+r_{\mu}))
  \Big]
\nonumber\\
&&+(3(\xM-1)+\xmu)L_2+(5\xM-3)(1-\xmu-L_2)\, ,
\eea
where $L_1\equiv \log[(1-\xmu)/\sqrt{\xmu x_e}]$, $L_2\equiv \log[(1-\xmu)/x_e]$.

The upper bounds (derived by the ${\rm BR}(\M\to \mu \mathnormal{e})$ upper limits   in Eqs.~(\ref{expboundsmeson})) for the 
$\emmp \to \M \gamma$ total cross sections  versus $\sqrt{s}/m_{\M}$ are reported in the right plot of Fig.~\ref{Fig3}. 
The infrared divergence of the cross-section near resonance is  tamed by requiring the detection of photons with energy 
$E_\gamma \gsim m_{\M}/10$~\cite{Karliner:2015tga}. 

Although suppressed by the $\alpha$ factor with respect to the resonant case, these cross sections retain values 
within experimental observability with presumably realistic running times, relaxing the stringent demand for small 
energy spreads as in the resonant case. 

It is also worth to point out that the effective Lagrangian approach,  
assuming pointlike and electrically neutral mesons, omits the emission of photons from the constituent quarks, the so-called 
structure-dependent emission, which is expected to contribute significantly at energies well above the threshold. 
Therefore our estimates at high energies should be considered as conservative, though in a regime which is anyway less favourable for the proposed experiments.

\section{Summary and conclusions}

The CLFV resonant productions considered in this analysis are  summarized in Table I, where we show, for each  process,  a few  relevant
quantities that are crucial to the actual experimental implementation. The kinematical constraint is expressed in terms of the  product of the 
beam energies to create the corresponding particle according to Eq.~(\ref{kinematic}), providing some flexibility in the choice of the colliding beams. 
The processes involving neutral mesons might be feasible once high-luminosity muon beams will be available. The corresponding electron 
beam does not have to be at comparable high energy, therefore benefiting from the existence of less expensive high-intensity electron accelerators 
with energy in the (1-100) MeV range.

Purely leptonic resonating processes we have considered appear much more challenging. There is a certain complementarity in this regard, 
and processes with higher cross-sections are unfortunately harder to achieve kinematically. The $\gamma e \to \mu$ process would be quite 
feasible in terms of kinematics, for instance by using high-intensity electron beams as the one used in synchrotron radiation machines, with 
energy $E_e\simeq$ 2.8 GeV. This would imply the use of high-intensity photon beams centered around 1~MeV, {\it i.e.} in the $\gamma$-ray range. 
This is not a trivial requirement and, in the absence of  high intensity $\gamma$-ray lasers, might be achieved only with high-intensity machines 
using inverse Compton scattering, but with unnaturally  small energy spread. There are already $\gamma$-ray facilities  which might be of some 
interest for this kind of experiments, such as ${\rm HI\gamma s}$~\cite{Weller} and ELI-NP~\cite{Filipescu}, or proposed ones like the Gamma factory at CERN~\cite{Budker}. One could alternatively use 
high-intensity photon beams in the visible region, with $E_\gamma \simeq 1~ \mathrm{eV}$, but only at 
the price of using electron beams with unrealistically high energy $E_e \simeq 2.8 \times 10^3$ TeV.  
The  $\gamma e \to \tau$and $\gamma \mu \to \tau$ cases have instead larger cross-sections, but the kinematics is less favourable, requiring high-energy muon beams with energy of order 1 TeV, and photon beams made of $\gamma$-rays in the 1~MeV range. Further analyses of the experimental feasibility of such collision setup will be required.

In conclusion, we have discussed the possibility of constraining lepton flavor violations in the charged sector using resonant and radiative-return  processes, corresponding to the inverse of presently explored decay modes. 
In particular,   we computed the upper  bounds on the resonant lepton ($\gamma\, \ell \to \ellp$) and neutral-meson 
($ \emmp \to \phi,\eta,\pi^0\!$) cross sections, and on the corresponding radiative processes. 
The characteristic of this approach is the possibility to control and 
boost the rates by proper engineering the  beam luminosity  setups. 
We have stressed the limitations due the the energy spread of the colliding beams, and discussed 
how to circumvent them by using the corresponding broadband radiative processes 
($\gamma\, \ell \to \gamma \ellp$ and $ \emmp \to \gamma\, (\phi,\eta,\pi^0)$), for which analytical cross sections
 have been computed by using an EFT approach. 

The proposal is more effective for particles with large total width, and in this sense might generate competitive bounds, especially in 
the mesonic processes. For the purely leptonic processes, it seems that a strong effort for achieving beams with smaller energy spread 
is necessary, making this a further demand to the ongoing research and development on muon accelerators and high flux gamma sources 
\cite{Ankenbrandt:1999cta,Gadjev:2017lvz,Ping,Raja:1998ip}.

From the present discussion is clear that the actual sensitivity of the considered channels to possible CLFV effects will crucially depend 
on  the luminosities and beams energy spread which can be actually reachable in the needed collision setup. On the other hand, here we 
did not focus on and discuss possible backgrounds of different nature, which in general will be as much crucial to set the actual potential 
of each channel and of  the corresponding experimental signature. We leave a detailed discussion of this issue to future dedicated studies.

\vfill

\appendix
\section{Differential $\gamma e \to \gamma \mu$ cross section}

Here we provide the exact expression for the differential cross section for the process $e(p_1) \gamma(q_1)\to \mu(p_2) \gamma(q_2)$ by
retaining all the mass dependences.  In particular we make explicit, in Eq.(\ref{dsigmarad}), the $F(s,t)$ function  as

\bea
F(s,t)&=&\frac{u \sum_{k=0}^{10}m_e^k F^{(k)}(s,t)}{D(s,t)}\, ,
\eea
with $D(s,t)= (s-m^2_e)^4 (t-m_e^2)^2(t-m_{\mu}^2)^2$, and $s,t$ are the 
Mandelstam variables as defined in Section III B, with $u=-s-t+m_e^2+\mmu^2$. 
The functions $F^{(k)}(s,t)$ are given by
\begin{widetext}
\bea
F^{(0)}(s,t)&=&-st\left(
6 s^3 t^3 - 12 \mmu^2 s^2 t^2 (s + t) +
\mmu^4 s t (7 s^2 + 24 s t + 7 t^2) -\mmu^6 (s + t) (s^2 + 13 s t + t^2)
+12 \mmu^8 s t-\mmu^{10} (s + t)\right)\, ,
\nonumber
\\
\nonumber
\\
\nonumber
F^{(1)}(s,t)&=&
2\mmu s^2 t^2 (s-\mmu^2 )  (t-\mmu^2 ) (s+t-\mmu^2)\, ,
\\
\nonumber
\\
\nonumber
F^{(2)}(s,t)&=&
12 s^3 t^3 (s + t) - 4 \mmu^2 s^2 t^2 (3 s + 2 t) (2 s + 3 t)
+ \mmu^4 s t (s + t) (13 s^2 + 55 s t + 13 t^2)\\ \nonumber
&-& 
   \mmu^6 (s^4 + 38 s^3 t + 50 s^2 t^2 + 38 s t^3 + t^4)
   + \mmu^8 (s + t) (2 s^2 + 21 s t + 2 t^2) - \mmu^{10} (3 s^2 + 2 s t + 3 t^2)\, ,
\\
\nonumber
\\
\nonumber
F^{(3)}(s,t)&=&
2\mmu st(s-\mmu^2 ) (t-\mmu^2) (\mmu^4 - s^2 - 3 s t - t^2)\, ,  
\\
\nonumber
\\
\nonumber
F^{(4)}(s,t)&=&
-s^2 t^2 (7 s^2 + 24 s t + 7 t^2)
  + \mmu^2 s t (s + t) (13 s^2 + 55 s t + 13 t^2) - 
   2 \mmu^4 (3 s^4 + 24 s^3 t + 76 s^2 t^2 + 24 s t^3 + 3 t^4)\\ \nonumber
&+& 2 \mmu^6 (s + t) (8 s^2 + 33 s t + 8 t^2) 
   - 7 \mmu^8 (s^2 + 8 s t + t^2) + 5 \mmu^{10} (s + t)
\\
\nonumber
\\
\nonumber
F^{(5)}(s,t)&=&
2 \mmu (s-\mmu^2 ) (t-\mmu^2) (s + t) (2 s t + 
\mmu^2 (s + t) -\mmu^4 )\, ,
\\
\nonumber
\\
\nonumber
F^{(6)}(s,t)&=&
s t (s + t) (s^2 + 13 s t + t^2)
-\mmu^2 (s^4 + 38 s^3 t + 50 s^2 t^2 + 38 s t^3 + t^4)
+ 2 \mmu^4 (s + t) (8 s^2 + 33 s t + 8 t^2)\\ \nonumber
&-& 6 \mmu^6 (7 s^2 + 8 s t + 7 t^2)   
+ 19 \mmu^8 (s + t) 
-4 \mmu^{10}\, ,
\\
\nonumber
\\
\nonumber
F^{(7)}(s,t)&=&
2 \mmu (s-\mmu^2 ) (t-\mmu^2) (\mmu^4 - s t - 2 \mmu^2 (s + t))\, ,
\\
\nonumber
\\
\nonumber
F^{(8)}(s,t)&=& - 12 s^2 t^2 
+ \mmu^2 (s + t) (2 s^2 + 21 s t + 2 t^2)
- 7 \mmu^4 (s^2 + 8 s t + t^2)+ 19 \mmu^6 (s + t)
-6 \mmu^8\, ,
\\
\nonumber
\\
\nonumber
F^{(9)}(s,t)&=&
2 \mmu^3 (s-\mmu^2) (t-\mmu^2 )\, ,
\\
\nonumber
\\
\nonumber
F^{(10)}(s,t)&=&
s t (s + t) - \mmu^2 (3 s^2 + 2 s t + 3 t^2) + 5 \mmu^4 (s + t)-4 \mmu^6\, .
\label{F}
\eea
\end{widetext}

We now provide an approximate formula for the total cross section, valid for the kinematical regions $s\sim m_{\mu}^2$ close to the resonance muon threshold.
For this aim, it is convenient to look at the angular distribution $\theta$, with $\theta$  the angle between the muon and electron momenta in the center-of-mass frame. In this case the variable $t$ can be expressed as
\bea
t=m_e^2-\frac{(s+m_e^2)(s-\mmu^2)}{2s}\left(1-\beta_e\cos{\theta}\right)\,,
\label{t}
\eea
with $\beta_e=(s-m_e^2)/(s+m_e^2)$ the electron velocity.
Then, the angular distribution for the cross section is given by
\bea
\frac{d \sigma}{d\cos{\theta}}(e \gamma\to \mu \gamma)&=&
\frac{\alpha  F(s,t)(1-r_e)}{4\Lambda_{\mu e}^2}\times  \nonumber\\
  &&\Big[\frac{s-\mmu^2}{
  (s-\mmu^2)^2+\Gamma_{\mu}^2\mmu^2}\Big]\,,
\label{dcost}
\eea
where $r_e=m^2_e/s$.
Notice that the $s$ distribution inside the square brackets in Eq. (\ref{dcost}) has a maximum for $s=\bar{s}$, where $\bar{s}\equiv \mmu^2(1+\delta)$ 
with $\delta\sim \Gamma_{\mu}/\mmu$, while the function $F(s,t)$ is almost flat in $s$ near regions close to $s \sim \bar{s}$. 
Therefore, in order to extract the dominant contribution to the total cross section relevant to the peak region, we approximate $F(s,t)$ with its expression evaluated at $s=\bar{s}$, thus by replacing  $F(s,t) \to F(\bar{s},\bar{t})$, where $\bar{t}\equiv t|_{s=\bar{s}}$, and averaged it over $-1<\cos{\theta}<1$.     
 In particular, by defining
\bea
\int_{-1}^{1} d\cos{\theta}~ \frac{F(\bar{s},\bar{t})}{4} \equiv C^{e \mu}_F \mmu^2 \, ,
\label{integral}
\eea
the approximated total cross section near the threshold is
\bea
\sigma_{\rm thr}(\gamma e \to \gamma \mu) &\simeq& \frac{\alpha \;C^{e\mu}_F}{\Lambda_{\mu e}^2}
\frac{\left(y_s-1\right)}{(y_s-1)^2+y^2_{\scriptscriptstyle{\Gamma_{\mu}}}}\, ,
\label{rad-cross-approx}
\eea
where $y_a\equiv a/\mmu^2$. By numerical integrating Eq.(\ref{integral}) we obtain $C^{e \mu}_F\simeq 8.7$.
The approximated formula in Eq. (\ref{rad-cross-approx}) fits with good accuracy the exact result, in particular, with an average of 20\% accuracy 
in the range $\mmu < \sqrt{s}< 2\mmu$ up to a few percent for $\sqrt{s}$ within 10\% from the resonant mass. 

For the analogous processes $\gamma e \to \gamma \tau$ and $\gamma \mu\to \gamma \tau$, the corresponding coefficients are $C^{e\tau}_F\simeq 14.3$ \; and\;  $C^{\mu\tau}_F\simeq 3.7$, respectively.


\end{document}